\documentclass[12pt]{article}
\newcommand{\Trg}{\mbox{Trg}\ }
\newcommand{\trg}{\mbox{trg}\ }
\newcommand{\Real}{\mbox{Re}}

\title{Random Matrix Theory by the Supersymmetry Method Beyond the
Sigma-Model}

\author{ Vladan Lu\v{c}i\'{c} 
\\
Physics Department, Northeastern University \\
Boston, MA 02115, USA 
}

\begin{document}

\maketitle

\begin{abstract}
The leading correction to the smoothed connected energy
density-density correlation function is obtained for the large energy
difference, within the context of the Gaussian Random Matrix Theory. In
order to achieve this result, the supersymmetry method is extended
beyond the sigma-model, to include small quadratic fluctuations around
the saddle point.  Special care is taken to avoid the potential
divergence arising from the unbounded nature of the saddle
point. Also, in the small energy difference regime, the leading
correction to a two point correlation function is obtained.
\end{abstract}


PACS-96: 05.40.+j

\newpage

\section{Introduction}

The Random Matrix Theory (RMT) has its roots in an idea of Wigner that
statistical mechanics should be used to analyze highly excited energy
levels of large nuclei \cite{Wignernuc}. This concept was further
developed by Dyson \cite{Dyson} to become what is now known as RMT.
Since then, it has been applied to numerous problems related to
complex quantum systems in different areas of physics \cite{Mehta}.

In RMT, the Hamiltonian $H$ is represented by a matrix of dimension $N$,
for $N$ large. The matrix is completely random, except that it
satisfies a certain symmetry. The ensemble average of any function
$f(H)$ of the Hamiltonian is calculated by multiplying the function by
a probability distribution and averaging over all possible matrices
that possess the required symmetry:
\begin{equation}
\langle f(H) \rangle = \frac{1}{Z} \int \mathcal{D} H f(H) 
\exp \{ -N Tr\left[ V(H)\right] \} ,
\label{eq:ansaverage}
\end{equation}
where $Z$ is the partition function:
\begin{equation}
Z = \int \mathcal{D} H \exp \{ -N Tr\left[ V(H)\right]
\}.
\end{equation}

RMT has been utilized for various purposes. For example, in condensed
matter physics of disordered systems, the Hamiltonian, or a part of
it, can often be taken to be random, except for some physical
symmetries.  Assuming that a particular physical quantity does not
depend on the details of the system, it can be calculated using
RMT. The problem of dealing with a complicated (or unknown)
Hamiltonian of the physical system is replaced in RMT by averaging
over an ensemble of random matrices which possess a certain symmetry,
as required by the symmetry of the original Hamiltonian. Commonly used
symmetries of the ensemble are unitary, orthogonal and symplectic. The
problems where RMT has been applied to include conductance
fluctuations in mesoscopic systems \cite{Iida}, persistent current of
disordered metal rings \cite{persist} and the motion of a quantum
particle in a thick wire \cite{Fyodorov}. A similar approach is taken
in analyzing the low energy behavior of QCD, only here the ensemble
considered possesses one of the chiral symmetries: chiral orthogonal,
chiral unitary, or chiral symplectic \cite{QCD}.

Another incentive for exploring RMT comes from its relation to both
classical and quantum chaotic systems.  On the basis of numerical
evidence, it was strongly conjectured that certain statistical
properties of quantum chaotic spectra are universal and identical to
those predicted by Gaussian ensembles of RMT \cite{Bohigas}. The
connection was further elaborated in \cite{SimLAlt} and proved in
semiclassical limit in \cite{AASA}.  Another
important problem in chaos, the transition from classical integrable
to chaotic behavior, has been investigated by relating classical chaos
to the size of a band of an ensemble of banded random matrices
\cite{clchaos}. Banded ensembles have been constructed to describe
the level fluctuations of a particular type of anharmonic oscillator
with a parameter that is responsible for the transition from
integrable to non-integrable behavior \cite{SelVZ}.

In all the applications of RMT mentioned above, one deals with
Gaussian ensembles. The potential $V$ in (\ref{eq:ansaverage}) has a
simple quadratic form (\ref{eq:GV}), which is equivalent to the
requirement that different elements of matrix $H$ be uncorrelated. It
is interesting to investigate how additional terms in the potential of
the non-Gaussian ensembles affect different correlators, or, in other
words, what are the quantities calculated within RMT that do not
depend on the form of the potential, but on the symmetries alone.
Universal properties of correlations for non-Gaussian ensembles and
their relation to the Gaussian ensembles were studied and the
correlations calculated, in the context of matrix models, in
\cite{loop} using the loop equations method. The same properties were
investigated in the context of disordered systems 
in \cite{BZ} using the orthogonal polynomials. The conjecture made
there was clarified in \cite{Eynard}, while the rigorous treatment was
given in \cite{FKY}. Other methods, such
as the diagrammatic method \cite{BZdiagram}, the functional derivative
method \cite{Beenakker} and the supersymmetry method \cite{nGsusy}
were also introduced and used for similar problems.

On the other hand, RMT can be extended to matrix models which have
been studied for their own sake. Matrix models, for example, has
been used to describe 2D gravity. The partition function of the
matrix models is obtained from the RMT partition function by adding a
kinetic term to the potential and summing 
over all Feynman diagrams, while in 2D gravity, the partition
function is obtained by summing over all surfaces. The potential term
generates Feynman diagrams, which can be regarded as the inverse
lattice of a discretized surface of arbitrary genus. More precisely, a
Feynman diagram is the inverse lattice of the discretized surface. The
ensemble average of RMT can now be regarded as a sum over all diagrams
(surfaces). Furthermore, when $N$ is large, the partition function has
a perturbative expansion in $N^{-1}$, each term corresponding to the
contribution of all surfaces of a fixed genus. It has been shown that
the dominant contribution comes from a sphere (genus 0)
\cite{'tHooft}, so the approximation in which only the leading term in
the expansion is used is called spherical or planar approximation.
The genus $1$ contribution to the partition function and the one point
correlator have been calculated in \cite{g1}, using the loop
equations. Good reviews of different aspects of matrix models are
given in \cite{mm}.

While the original approach used in RMT was the method of orthogonal
polynomials \cite{Mehta}, the supersymmetry method, pioneered by
Efetov \cite{Efetov}, has been also proven to be very useful,
especially when the coupling to external sources is needed
\cite{Iida}. It is essentially an extension of Anderson's replica
trick \cite{VZ}.  A very thorough and detailed review of the method can
be found in \cite{VWZ}.

The results cited above were concerned with the leading order
contributions to different RMT correlators. The leading order
correction (genus one contribution, in the language of matrix models)
to the one-point correlation function for the Hermitian matrix model has
been calculated in \cite{genus1}, while the iterative procedures to
calculate other corrections (higher geni contributions) were presented
for Hermitian \cite{iterativeHermitian}, complex
\cite{iterativeComplex}, and orthogonal and symplectic
\cite{iterativeOS} matrix models. Recently, there has been significant
interest in finding the corrections within the context of disordered
systems. Orthogonal polynomials have been
used to calculate leading corrections to one and two-point correlation
functions for the complex RMT \cite{correctionComplex}. The supersymmetry
method has also been used to calculate corrections to various
quantities within RMT: the energy-energy correlation in the regime where
the energy difference is small \cite{2ptCorrection}, the level curvature
distribution \cite{levelCurvature} and the eigenfunction amplitudes
distribution \cite{eigenfunction}.

In this paper, the leading order correction to two-point correlators,
in the $N^{-1}$ expansion, is calculated using the extension of the
supersymmetry method.  Within the usual supersymmetry method, the
partition function is expanded around the saddle point (the saddle
point manifold, to be precise), the energy difference is considered
small ($\sim N^{-1}$), and only the contribution coming from the
saddle point is considered. This simplification leads to the so-called
zero-dimensional non-linear $\sigma$-model. Here, the method is
extended beyond the $\sigma$-model: both to accommodate for the large
energy difference ($\sim N^0$) and to include small fluctuations
around the saddle point. By extending the method for the large energy
difference the leading contribution to the two point correlators can
be obtained (in the large energy regime), while the inclusion of the
fluctuations allows the calculation of the corrections in both
regimes.  As it is customary in the large energy regime, the smoothed
energy density-density correlation function is considered \cite{BZ};
that is, the correlator is averaged over the energy scale of the order
of few $N^{-1}$'s, so that the terms oscillating with the frequency of
the order $N$ go to zero. The leading order contribution of the
(smoothed) connected energy density-density correlation function
obtained in this text agrees with the result obtained using the method
of orthogonal polynomials, \cite{BZ}.  In the small energy difference
regime, the correction to the connected two point correlator, $\langle
G^+(E_1) G^-(E_2) \rangle^c$, is obtained. The leading term of the
correlator is also calculated and it is found to agree with the well
known result which was obtained using the supersymmetry, as well as
other methods.

Furthermore, the convergence of the method is proved. Particular care
has to be taken because of the potential divergence arising from the
unbounded nature of the saddle point manifold. The singularity
structure of Green's functions is used in order to resolve the
ambiguities arising from the double-valued nature of the saddle point.
As it is usual in the supersymmetry method, the Gaussian ensemble is
considered in the limit when $N \rightarrow \infty$ and the energies
are taken not too close to the ends of the spectrum. For the sake of
the definiteness, in this text, the ensemble with the unitary symmetry
is considered.

The approach used in this text was to follow the presentation of the
supersymmetry method as given in an excellent introductory paper by
Zuk \cite{Zuk} and to modify it when needed. Some parts of the
calculations were done using the symbolic software package Dill
\cite{Dill} as well as additional programs written in the
Mathematica${}^{\copyright}$ programming language \cite{Mathematica}.

The paper is organized as follows: the important steps of the usual
supersymmetry formalism are given in section \ref{sec:supm}. The
definitions and conventions related to the Grassman variables and the
integration over the Grassman variables are given in appendices
\ref{ap:Gdef} and \ref{ap:Gint}, respectively.  The formalism is
modified to include the corrections in section \ref{sec:bes}, while
the correlation functions are calculated and the convergence of the
method is proved in section \ref{sec:cor},
using the results presented in appendix \ref{ap:GGc}.

\section{Short Review of the Supersymmetry Method}
\label{sec:supm}

In this section, the main points of the supersymmetry method are
summarized. This is needed in order to show how the extended
supersymmetry method is constructed and to prove its convergence. We
begin by defining the relevant quantities.  The quantity of main
interest here, the connected two point correlation function, is given
by:
\begin{equation}
\langle {G^+(E_1) G^{\pm}(E_2)} \rangle^c =
\langle {G^+(E_1) G^{\pm}(E_2)} \rangle - 
\langle {G^+(E_1)} \rangle \langle {G^{\pm}(E_2)} \rangle,
\label{eq:Gcdef}
\end{equation}
where:
\begin{equation}
G^{\pm}(E) = \mbox{Tr} {1 \over E - H \pm i\, \epsilon},
\label{eq:Gdef}
\end{equation}
and the averaging $\langle \ldots \rangle$ is performed over Gaussian
unitary ensemble; therefore, it is given by (\ref{eq:ansaverage})
with the potential $V$ given by:
\begin{equation}
V(H) = \frac{1}{2 \lambda^2}\, H^2,
\label{eq:GV}
\end{equation}
where $\lambda$ is an arbitrary constant. 

The density of states is given by:
\begin{equation}
\rho = {i \over 2\, \pi\, N} \left[ G^+(E) - G^-(E) \right], 
\end{equation}
so that the connected part of the ensemble averaged energy
density-density correlation function is given by:
\begin{eqnarray}
C(E_1, E_2) & = & \langle {\rho(E_1)\, \rho(E_2)} \rangle - 
\langle {\rho(E_1)} \rangle \, \langle {\rho(E_2)} \rangle
\nonumber \\
& = &
\frac{1}{2 \pi^2 N^2} \mbox{Re} \left[
\langle {G^+(E_1)G^-(E_2)} \rangle^c -
\langle {G^+(E_1)G^+(E_2)} \rangle^c \right].
\label{eq:CbyGG}
\end{eqnarray}  

The density of states satisfies Wigner's semi-circle law
\cite{Wigner}:
\begin{equation}
\rho(E) = \frac{1}{2 \pi \lambda^2} \sqrt{4 \lambda^2 - E^2},
\end{equation}
so that the $N$ energy eigenstates of $H$ are confined to a region between $2
\lambda$ and $-2 \lambda$.

In order to begin with the supersymmetry formalism, let us consider
now a generating function:
\begin{eqnarray}
Z(\varepsilon) & = & \mbox{Detg}^{-1}\left[D + J(\varepsilon)\right]
\nonumber \\
& = & \exp \{-\Trg \ln \left[D + J(\varepsilon)\right]\}
\label{eq:Zbydet}
\end{eqnarray}
where Detg and Trg denote the graded (super) determinant and the
graded (super) trace respectively \cite{Nath} and are taken over all
indices. The symbol trg is used when the summation is to be taken over
$p$ and $\alpha$ indices only.  (For a complete introduction to
anti-commutative (Grassman) variables one can consult \cite{Berezin}
and \cite{VWZ}.)  The definitions and conventions used in this article
related to the Grassman variables are given in appendix
\ref{ap:Gdef}. $D$ and $J$ are $4N \times 4N$ graded matrices,
diagonal in both upper and lower indices, the only non-vanishing
elements being:
\begin{eqnarray}
D^{00}_{pp} =  D^{11}_{pp} & = & (E_1 + i \epsilon_p)\, \mathbf{1}_N - H,
\\ 
J^{\alpha \alpha}_{p p}(\varepsilon) & = & (-)^{\alpha}
\varepsilon_p \mathbf{1}_N.
\label{eq:Jdef} 
\end{eqnarray}

It is then easy to show that:
\begin{eqnarray}
G(E_p)  =  -\frac{1}{2} \left. \frac{ \partial }{\partial \varepsilon_p}
 Z(\varepsilon)\right|_{\varepsilon = 0} 
\makebox[60pt]{} \label{eq:GbydZ} \\
G(E_1 + i \epsilon_1) G(E_2 + i \epsilon_2) =
\frac{1}{4} \left. \frac{ \partial^2 }{\partial \varepsilon_1 \partial
\varepsilon_2}
 Z(\varepsilon) \right|_{\varepsilon = 0}
\label{eq:GGbydZ}
\end{eqnarray}
The two point Green's functions, $G^+(E_1) G^-(E_2)$ and $G^+(E_1)
G^+(E_2)$, are obtained from the above equation by taking
$\epsilon_1 = \epsilon = -\epsilon_2$ and $\epsilon_1 = \epsilon =
\epsilon_2$ respectively, for $\epsilon$ positive and infinitesimal.
Therefore $Z(\varepsilon)$ is indeed the generating function for the
two point Green's functions. In this case, the matrix $D + J$ consists
of two diagonal blocks, corresponding to the two Green's
functions. The two blocks contain the information about the respective
energies as well as the sides of the branch cut of the Green's
function from which the energy approaches the real axis. If the
ensemble averaged product of more Green's functions were to be
calculated, additional blocks would have to be added to $D + J$.

Some important properties of the integration over Grassman variables
are given in appendix \ref{ap:Gint}.  Using the Gaussian
super-integral (\ref{eq:superGauss}), the generating function
(\ref{eq:Zbydet}), can be ensemble averaged and shown to be:
\begin{eqnarray}
\langle Z(\varepsilon) \rangle =
\int \mathcal{D} \phi \mathcal{D} {\bar{\phi}} \;
\langle{\exp \left[ -i \sum_{p, \alpha, \mu} \bar{\phi}_p^\alpha
(\mu) H \phi_p^\alpha (\mu) \right] }\rangle
\times \makebox[100pt] \nonumber \\
\exp\; i \left[ \sum_{p, \alpha, \mu} \bar{\phi}_p^\alpha (\mu) 
(E + \frac{\omega}{2} L + i \epsilon_1 I_1 + i \epsilon_2 I_2 + J)  
\phi_p^\alpha (\mu) \right],
\label{eq:L1}
\end{eqnarray}
where the average energy $E = (E_1 + E_2) / 2$, and the difference
between the energies $\omega = E_1 - E_2$ have been introduced and
$L$, $I_1$ and $I_2$ are diagonal in $p$ and $\alpha$ indices:
\begin{equation}
L^{\alpha \alpha^\prime}_{p p^\prime} = 
(-)^{p - 1} \delta^{\alpha \alpha^\prime} \delta_{p p^\prime},\qquad
(I_i)^{\alpha \alpha^\prime}_{p p^\prime} = 
\delta^{\alpha \alpha^\prime} \delta_{p p^\prime} \delta_p^i.
\end{equation}

At this point, it is clear that due to the Grassman variables, the
generating function is normalized by the definition (\ref{eq:Zbydet})
($Z(0) = 1$) which makes it possible to interchange the integration
over $\phi$'s and the ensemble averaging (integration over the random
matrix $H$) and, eventually, to perform the ensemble averaging
completely. In the case where a generating function is not normalized,
it has to be normalized first by dividing by the corresponding
partition function and then ensemble averaged. Typically, both the
partition function and the generating function are integrals, so it is
a very nontrivial task to perform the ensemble averaging.

The procedure is continued using:
\begin{equation}
\langle {\exp \left[ -i \sum_{p, \alpha, \mu} \bar{\phi}_p^\alpha
(\mu) H \phi_p^\alpha (\mu) \right] }\rangle = 
\exp -\frac{\lambda^2}{2 N} \trg S^2, \qquad
\mbox{\hfill}
S^{\alpha \alpha^\prime}_{p p^\prime} = 
\sum_{\mu} \phi^{\alpha}_p (\mu) \bar{\phi}^{\alpha^\prime}_{p^\prime}(\mu). 
\label{eq:Sdef}
\end{equation}
Then the ensemble average over the random Hamiltonian can be completely
performed to give:
\begin{equation}
\langle Z(\varepsilon) \rangle =
\int \mathcal{D} \phi \bar{\phi} 
\exp \left[ -\frac{\lambda^2}{2 N} \trg S^2 +
i \trg (E + \frac{\omega}{2} L + i \epsilon_1 I_1 + i \epsilon_2 I_2 +
J) S \right].
\label{eq:L2}
\end{equation}

Obviously, the price that has to be paid for carrying out the ensemble
averaging completely is the introduction of the quartic terms ($S^2$)
in the integrand. The way to proceed is to eliminate these terms, using the 
Hubbard-Stratonovich transformation, as follows.

The function $W$ is defined by:
\begin{equation}
i W(\sigma, S) = 
\frac{\lambda^2}{2 N} \trg (S - \frac{i N}{\lambda} \sigma)^2 + 
\trg ( \epsilon_1 I_1 + \epsilon_2 I_2 )
(S - \frac{i N}{\lambda} \sigma),
\label{eq:Wdef}
\end{equation}
where $\sigma$ is an arbitrary $4 \times 4$ super-matrix having the
same symmetry properties as $S$. Using the parametrization for
$\sigma$ introduced below, and as long as $N$ is large, it follows that:
\begin{equation}
\int \mathcal{D} \sigma e^{i W(\sigma, S)} =
\int \mathcal{D} \sigma e^{i W(\sigma, 0)} = 1.
\label{eq:expW}
\end{equation}
The second identity is easy to check once the parametrization for
$\sigma$ is introduced (\ref{eq:sigmadec})--(\ref{eq:Ds}).
After the previous identity is inserted in (\ref{eq:L2}) and the
shift of variables:
\mbox{$\sigma \mapsto \sigma - \frac{1}{\lambda}(\frac{\omega}{2} L + J)$}
is performed, the averaged generating function takes the form:
\begin{eqnarray}
\langle Z(\varepsilon) \rangle = 
\int \mathcal{D} \sigma \exp \; \left\{
-\frac{N}{2} \trg \sigma^2 - 
\frac{N \omega}{2 \lambda} \trg \sigma L -  
\frac{i N}{\lambda} \trg (\epsilon_1 I_1 + \epsilon_2 I_2) \sigma - 
\right.  \label{eq:L3}
\\
N \trg \ln (E - \lambda \sigma ) - 
\left. \frac{N}{\lambda} \trg \left[
\sigma + \frac{\omega}{2 \lambda} L +
\frac{i}{\lambda} (\epsilon_1 I_1 + \epsilon_2 I_2)
\right] J \right\}
\nonumber
\end{eqnarray}
completing the Hubbard-Stratonovich transformation.

At this point, it is important to notice that not only have the
quartic terms disappeared from the generating functional, but in
addition the number of integration variables have decreased from $4 N$
complex variables in (\ref{eq:L1}) to those spanned by the $4 \times
4$ complex super-matrix $\sigma$ in (\ref{eq:L3}). Not all entries
in $\sigma$ are independent, because it has to satisfy certain
symmetry properties. Therefore, in order to see what exactly the
integration variables are, the symmetries of the super-matrix $\sigma$
have to be investigated first and then the parametrization of $\sigma$
(the parameters being the independent variables) consistent with the
symmetries can be found. The detailed account of the symmetries and
the parametrization leading to the independent variables of the
super-matrix $\sigma$ for the Gaussian orthogonal ensemble is given in
\cite{VWZ}. The analysis for the case needed here, the Gaussian
unitary ensemble, is similar and is clearly stated in \cite{Zuk}; what
follows is only a short description of the results.

In order to satisfy (\ref{eq:expW}), and having in mind the definition
of $W$ given in (\ref{eq:Wdef}), the super-matrices $\sigma$ and $S$
have to have the same symmetry. From the definition of $S$, (\ref{eq:Sdef})
it follows that: $S^{\dagger} = s S s$, so: $\sigma^\dagger = s \sigma
s$ also, leading to the decomposition:
\begin{equation}
\sigma = T^{-1} P_d T,
\label{eq:sigmaform}
\end{equation}
where $4 \times 4$ super-matrix $T$ has to satisfy:
\begin{equation}
T^{-1} = s T^{\dagger} s,
\label{eq:sTs}
\end{equation}
and $P_d$ has to be a diagonal super-matrix with real boson-boson and
imaginary fermion-fermion elements. The last condition has to be
changed slightly in order to achieve convergence of integrals in
(\ref{eq:expW}) and make $P_d$ belong to the saddle point of the exponent
in (\ref{eq:L3}). Instead of (\ref{eq:sigmaform}), $\sigma$ is
decomposed as:
\begin{equation}
\sigma = T^{-1} (\sigma_0 + N^{-1/2} P_d) T.
\end{equation}
The change amounts to a simple shift of the diagonal matrix $P_d$ by a
constant diagonal $4 \times 4$ super-matrix $\sigma_0$ determined
below. $P_d$ now has real boson-boson and imaginary fermion-fermion
elements. Also, $P_d$ has been rescaled for future convenience.

Due to its defining property (\ref{eq:sTs}), the boson-boson part
of the super-matrix $T$ spans the (non-compact) group $SU(1,1)$ while
the fermion-fermion part spans the (compact) group SU(2). Further, $T$
can be written as $T = R T_0$, where $[R, L] = 0$, so that $R$ is
block-diagonal in the lower indices, and the blocks are denoted by
$R_1$ and $R_2$. The symmetry of the matrix $T$ implies: $T^{-1}_0 = s
T^{\dagger}_0 s$, $R^{-1}_1 = R^{\dagger}_1$ and $R^{-1}_2 = k
R^{\dagger}_2 k$.  This gives the final decomposition of $\sigma$:
\begin{equation}
\sigma = T^{-1}_0 (\sigma_0 + N^{-1/2} \delta P) T_0 
= {T^{-1}}_0 P {T}_0,
\label{eq:sigmadec}
\end{equation}
where
$
\delta P = R^{-1} P_d R.
$
Thus $\delta P$ is a block-diagonal super-matrix with real boson-boson and
imaginary fermion-fermion eigenvalues. The decomposition of $\sigma$ 
(\ref{eq:sigmadec}), can be regarded as splitting of the manifold
spanned by the independent variables of $\sigma$ to the space spanned by
$\delta P$ and the coset-manifold defined by $T_0$. Due to the change of
variables, the original measure $\mathcal{D} \sigma$ becomes:
\begin{equation}
\mathcal{D} \sigma = I(P) \mathcal{D} P \mathcal{D} \mu (T_0),
\label{eq:Dsigma}
\end{equation}
where $\mathcal{D} \mu (T_0)$ is the invariant measure on the coset
manifold and $I(P)$ is the Jacobian of the transformation $I(P)$:
\begin{equation}
I(P) = {\left[ \frac
{(\Lambda^0_1 - \Lambda^0_2) (\Lambda^1_1 - \Lambda^1_2)}
{(\Lambda^0_1 - \Lambda^1_2) (\Lambda^1_1 - \Lambda^0_2)} 
\right]}^2,
\label{eq:ipdef}
\end{equation}
and ${\Lambda^\alpha}_p$ are the eigenvalues of $P_p$. The derivation of
the last expression is straightforward; useful hints are
given in similar derivations of \cite{Wegth} and \cite{VWZ}.

Following its symmetry, the matrix $T_0$ can be decomposed as:
\begin{equation}
T_0 = U^{-1} T_d U,
\end{equation}
where:
\begin{equation}
U = \left( \begin{array}{cc}
u & 0 \\
0 & v \end{array} \right), \qquad
u = \exp \left( \begin{array}{cc}
		0 & -\eta^* \\
		\eta & 0 \end{array} \right), \qquad
v = \exp \left( \begin{array}{cc}
		0 & -i \rho^* \\
		i \rho & 0 \end{array} \right),
\label{eq:T0}
\end{equation}
$\eta$ and $\rho$ are Grassman variables, and
\begin{equation}
T_d = \left( \begin{array}{cccc}
\sqrt{1 + \mu_0 \bar{\mu}_0} & 0 & \mu_0 & 0 \\
0 & \sqrt{1 + \mu_1 \bar{\mu}_1} & 0 & \mu_1 \\
\bar{\mu}_0 & 0 & \sqrt{1 + \mu_0 \bar{\mu}_0} & 0 \\
0 & \bar{\mu}_1 & 0 & \sqrt{1 + \mu_1 \bar{\mu}_1}  \end{array}
\right)
\label{eq:Td}
\end{equation}
\begin{eqnarray}
\mu_0 = -e^{i \phi_0} \sinh \frac{\theta_0}{2}, & 
\bar{\mu}_0 = -e^{-i \phi_0} \sinh \frac{\theta_0}{2} \nonumber \\
\mu_1 = i e^{i \phi_1} \sin \frac{\theta_1}{2}, & 
\bar{\mu}_1 = i e^{-i \phi_1} \sin \frac{\theta_1}{2}. 
\end{eqnarray}
$\phi_0$, $\phi_1$ and $\theta_1$ are bounded (from $0$ to $2
\pi$), while $\theta_0$ is unbounded.

It is also useful to introduce:
\begin{eqnarray}
\lambda_0 \equiv \cosh \theta_0 = 
2 \mu_0 \bar{\mu}_0 + 1, & 1 \leq \lambda_0 < \infty 
\nonumber \\
\lambda_1 \equiv \cos \theta_1 =
2 \mu_1 \bar{\mu}_1 + 1, & -1 \leq \lambda_1 \leq 1.
\label{eq:lambdadef}
\end{eqnarray}
With the change of variables stated above, the invariant measure on
the coset manifold is found to be:
\begin{equation}
\mathcal{D} \mu(T_0) = \frac{1}{(\lambda_0 - \lambda_1)^2} 
d \lambda_0 d \lambda_1 d \phi_0 d \phi_1 
d \eta d \eta^* d \rho d \rho^*.
\label{eq:DT0}
\end{equation}

It is obvious from the previous equations that the symmetries of the
super-matrix $T_d$ are $SU(1,1) / U(1)$ for the boson-boson part and
$SU(2) / U(1)$ for the fermion-fermion part. Some parameters of the
original $SU(1,1) \times SU(2)$ of $T$ have been absorbed in the
degrees of freedom other than those of $T_d$, nevertheless the
non-compact degree of freedom can be identified as $\lambda_0$.

Finally, due to the symmetries of its factors, $\delta P$ is decomposed to:
\begin{equation}
\delta P = \left( \begin{array}{cc}
		\delta P_1 & 0 \\
		0 & \delta P_2 \end{array} \right),
\label{eq:Pdef}
\end{equation}
where:
\begin{eqnarray}
\delta P_1 & = \exp 
\left( \begin{array}{cc} 0 & \zeta^*_1 \\
-\zeta_1 & 0 \end{array} \right)
\ \left( \begin{array}{cc} a_1 & 0 \\
0 & i b_1 \end{array} \right)
\ \exp \left( \begin{array}{cc} 0 & -\zeta^*_1 \\
\zeta_1 & 0 \end{array} \right)
\nonumber \\
\delta P_2 & = \exp 
\left( \begin{array}{cc} 0 & i \zeta^*_2 \\
-i \zeta_2 & 0 \end{array} \right)
\ \left( \begin{array}{cc} a_2 & 0 \\
0 & i b_2 \end{array} \right)
\ \exp \left( \begin{array}{cc} 0 & -i \zeta^*_2 \\
i \zeta_2 & 0 \end{array} \right)
\label{eq:P12def}
\end{eqnarray}
$\zeta_1$ and $\zeta_2$ being Grassman variables and $a_1$, $a_2$,
$b_1$ and $b_2$ ranging from $-\infty$ to $\infty$.
The measure on $\delta P$ is easily computed to be:
\begin{equation}
\mathcal{D} P = \frac{-1}{(a_1 - i b_1)^2 (a_2 - i b_2)^2}
da_1da_2db_1db_2 d\zeta_1 d{\zeta^*}_1 d\zeta_2 d{\zeta^*}_2.
\label{eq:DP}
\end{equation}
Using the above stated parametrization for $\delta P$ and the
definition of $I(P)$, (\ref{eq:ipdef}) it follows that:
\begin{equation}
I(P) =
{\left[ \frac
{(\sigma_{01} - \sigma_{02} + N^{-1/2} (a_1 -  a_2)) 
(\sigma_{01} - \sigma_{02} + N^{-1/2} i (b_1 - b_2))}
{(\sigma_{01} - \sigma_{02} + N^{-1/2} (a_1 - i b_2)) 
(\sigma_{01} - \sigma_{02} + N^{-1/2} (i b_1 - a_2))} 
\right]}^2,
\label{eq:ip}
\end{equation}
where the $\sigma_{0p}$ are the saddle points determined in the next
section.  Combining (\ref{eq:Dsigma}), (\ref{eq:DT0}) and (\ref{eq:DP}),
the final expression for the measure on the $\sigma$-manifold is given
by:
\begin{equation}
\mathcal{D}\sigma =
-I(P) 
\frac{d \lambda_0 d \lambda_1 d \phi_0 d \phi_1 
d \eta d \eta^* d \rho d \rho^*
da_1da_2db_1db_2 d\zeta_1 d{\zeta^*}_1 d\zeta_2 d{\zeta^*}_2}
{(\lambda_0 - \lambda_1)^2 (a_1 - i b_1)^2 (a_2 - i b_2)^2},
\label{eq:Ds}
\end{equation}
with $I(P)$ given above (\ref{eq:ip}).

Therefore, an exact integral expression for the ensemble averaged
partition function (\ref{eq:L3}) has been found in this section.  Also
determined is the parametrization of the super-matrix $\sigma$ in
terms of the independent integration variables (\ref{eq:sigmadec}),
(\ref{eq:T0}), (\ref{eq:Td}), (\ref{eq:lambdadef}), (\ref{eq:Pdef})
and (\ref{eq:P12def}), and the expression for the measure in terms of
the integration variables (\ref{eq:Ds}). The connected part of the
ensemble averaged density-density correlation function, the quantity
of the main interest in this article, is determined from the
derivatives of the averaged partition function, as follows from
(\ref{eq:CbyGG}) and (\ref{eq:GGbydZ}).

\section{Supersymmetry Formalism Beyond the $\sigma$-model}
\label{sec:bes}

In this section, the calculation of the averaged partition function is
continued using the saddle point method. Contrary to the usual
supersymmetry method, where $\omega$ is taken to be small (of the
order of $N^{-1}$) and only the highest order terms in $N$ are
retained, both the leading and some sub-leading order terms are kept
in this article. In other words, in addition to the contribution
coming from the saddle point as in the standard supersymmetry method,
the small quadratic fluctuations, massive modes, are also considered here.

The saddle point of the exponent in (\ref{eq:L3}) can be obtained as
a solution of the following equation:
\begin{equation}
\frac{\delta}{\delta \sigma} \left[ \frac{1}{2} \trg \sigma^2 +
\frac{\omega}{2 \lambda} \trg \sigma L + 
\frac{i}{\lambda} \trg (\epsilon_1 I_1 + \epsilon_2 I_2) \sigma + 
\trg \ln (E - \lambda \sigma) \right] = 0.
\label{eq:saddlepoint}
\end{equation}
The solution of the previous equation is not unique. If the saddle
point is required to be diagonal, the diagonal elements $\sigma_{0p}$
have to satisfy the following equation:
\begin{equation}
\sigma_{0p} - \frac{\lambda}{E - \lambda \sigma_{0p}} +
(-)^{p-1} \frac{\omega}{2 \lambda} + 
(-)^{p-1} \frac{i \epsilon_p}{\lambda} =
0.
\label{eq:sigma0eq}
\end{equation}
The solution, for both values of $p$, is given by:
\begin{equation}
\sigma_{0p} = E^\prime + (-)^p \omega^\prime - 2 i {\epsilon^\prime}_p 
\pm i \sqrt{1 - 
\left[E^\prime - (-)^p \omega^\prime + 2 i {\epsilon^\prime}_p \right]^2},
\label{eq:sigma0p}
\end{equation}
where $E$, $\omega$ and $\epsilon_p$ has been rescaled:
\begin{equation}
E^{\prime} = \frac{E}{2 \lambda}, \quad
\omega^\prime = \frac{\omega}{4 \lambda} \quad \mbox{and} \quad
\epsilon^{\prime}_p = \frac{\epsilon_p}{2 \lambda}.
\end{equation}

$\sigma_{0p}$, ($p = 1, 2$) can be regarded as the two factors of the
correlator $\langle G(E_1) G(E_2)\rangle $. Furthermore, for
$\epsilon_p$ positive or negative in (\ref{eq:sigma0p}), $ G^+(E_p)$
or $G^-(E_p)$ respectively, are obtained, as shown below
(\ref{eq:GGbydZ}). The Green's function $G(z)$ as defined by
(\ref{eq:Gdef}) ($z$ complex), has a branch cut for $z$ real and $-2
\lambda \leq z \leq 2 \lambda$. That means that the points
\mbox{$G(E+i \epsilon) \equiv G^+(E) $} and \mbox{$ G(E-i \epsilon)
\equiv G^-(E)$}, for \mbox{$-2 \lambda \leq E \leq 2 \lambda$}, belong
to different Rieman surfaces, so that the sign of the small imaginary
part $i \epsilon$ determines the correct surface.  On the other hand,
the square in (\ref{eq:sigma0p}) also has a branch cut. Furthermore,
it is again the sign of the $\epsilon^{\prime}_p$ under the square
root that determines the sign of the imaginary part of the square root
and consequently the Rieman surface. Therefore, the branch cut
structure of the Green's function arises from the branch cut of the
(square root in the) expression for the saddle point $\sigma_{0p}$. As
in the standard $\sigma$-model, the convergence can only be achieved
if the $\pm$ sign in front of the square root corresponds to
$G^{\mp}(E_p)$ \cite{VWZ}. The convergence of the extended model,
presented here, is proved in the next section. Therefore, the
solutions of (\ref{eq:sigma0eq}) are:
\begin{eqnarray}
\sigma_{01}^{+\mp} & = & E^\prime - \omega^\prime -
\epsilon^\prime
\frac{E^\prime + \omega^\prime}{\sqrt{1 - (E^\prime + \omega^\prime)^2}} -
i \sqrt{1 - (E^\prime + \omega^\prime)^2} - i \epsilon^\prime
\nonumber \\
\sigma_{02}^{+\mp} & = & E^\prime + \omega^\prime -
\epsilon^\prime
\frac{E^\prime - \omega^\prime}{\sqrt{1 - (E^\prime - \omega^\prime)^2}} \pm
i \sqrt{1 - (E^\prime - \omega^\prime)^2} \pm i \epsilon^\prime
\label{eq:sigma0} 
\end{eqnarray}
Eventually, $\epsilon$ (in the equations above) will be set to zero,
but the convergence properties have to be determined before. Further, the
diagonal saddle point is defined as:
\begin{equation}
\sigma_0 = \mbox{diag} (\sigma_{01},\ \sigma_{01},\ \sigma_{02},
\ \sigma_{02}).
\label{eq:sigma0mat}
\end{equation}

In the standard supersymmetry method, $\omega$ is taken to be small,
so that the second term in the equation for the saddle point,
(\ref{eq:saddlepoint}), is neglected.  In that case, a matrix of the
form $T^{-1}_0 \sigma_0 T_0$ also satisfies the saddle point
condition. This is because the first and the last
terms are obviously unchanged by the transformation $\sigma_0 \mapsto
T^{-1}_0 \sigma_0 T_0$ and the third is small because it is
proportional to $\epsilon$. Consequently, the saddle point is actually
a manifold, parametrized by the degrees of freedom of matrix $T_0$.

The approach taken in this paper is that $\omega$ is not
neglected. Using the parametrization introduced in the previous
section: (\ref{eq:T0}), (\ref{eq:Td}) and (\ref{eq:lambdadef}), the
transformation of the second term of the saddle point equation
(\ref{eq:saddlepoint}) can be calculated:
\begin{equation}
\trg T^{-1}_0 \sigma_0 T_0 L = 
(\sigma_{01} - \sigma_{02}) (\lambda_0 - \lambda_1) \neq
\trg \sigma_0 L.
\end{equation}
This lifts the degeneracy of the saddle point. If $\omega$ is
small, $\omega \sim N^{-1}$, the manifold spanned by the matrix
$\sigma$ is almost flat in the direction of the matrix $T_0$ compared
to the directions of the other degrees of freedom (matrix $\delta
P$). Consequently, the manifold $T^{-1}_0 \sigma_0 T_0$ can still be
considered to be a saddle point manifold. Therefore it makes sense to
integrate over it (over the almost flat direction of the matrix
$T_0$), provided that the following constraint is observed. Since
$\lambda_0$, one of the degrees of freedom of the matrix $T_0$, is
unbounded (\ref{eq:T0})--(\ref{eq:lambdadef}), its upper bound should
be small compared to $N$ in order that the second, degeneracy lifting,
term of the saddle point equation (\ref{eq:saddlepoint}) stays
small compared to the other terms. The upper bound of $\lambda_0$ is
still taken to be infinity, provided that in the
limiting procedure $N \rightarrow \infty$ first and only then the
limit of the upper bond of $\lambda_0$ is taken.

In the large $\omega$ regime, the degeneracy of the saddle point
manifold is lifted fully, so that the degrees of freedom of both
$T_0$ and $\delta P$ can be considered as fluctuations around the
saddle point.

The decomposition of the matrix $\sigma$ given in
(\ref{eq:sigmadec}) should now be much more transparent: it is simply
the expansion around a saddle point. $N^{-1/2} T^{-1}_0 \delta P T_0$
are the fluctuations around the saddle point and $T_0$ represents the
saddle point manifold or the fluctuations, depending on the
regime. The factor $N^{-1/2}$ is introduced because it is customary in
the saddle point method to suppress the fluctuations by the inverse
square root of the coupling constant \cite{Ramond}. In effect, it
gives another limiting procedure -- the infinite bounds of the bosonic
degrees of freedom of the matrix $\delta P$ are of the order
$N^{1/2}$.

Both $\langle {G^+(E_1) G^+(E_2)} \rangle $ and $\langle {G^+(E_1)
G^-(E_2)} \rangle $ are calculated from $\langle Z(\varepsilon)
\rangle$; the difference lies in the sign of $\epsilon_2$ and in the
saddle points (\ref{eq:sigma0}), used in the expression for
$\langle Z(\varepsilon) \rangle$. Neither $\epsilon_1$ nor
$\epsilon_2$ play any role in the further calculations related to
$\langle {G^+(E_1) G^+(E_2)} \rangle $, so they can be set to zero.
However, they are still needed for $\langle {G^+(E_1) G^-(E_2)} \rangle
$. Both facts can be expressed using ${\omega^+}^\prime$ with the
understanding that:
\begin{equation}
{\omega^+}^\prime = \left\{ 
\begin{array}{cl} 
\omega^\prime + i\epsilon^\prime & \mbox{for the calculation of}  
\quad \langle {G^+(E_1) G^-(E_2)} \rangle
\\
\omega^\prime & \mbox{for the calculation of} 
\quad \langle {G^+(E_1) G^+(E_2)} \rangle
\end{array} \right.
\end{equation}

Expanding the exponent of the averaged partition function,
(\ref{eq:L3}), around the saddle point (\ref{eq:sigmadec}), keeping
the terms to the order of $N^0$ in the expansion of the logarithm, and
bearing in mind the definitions for $\sigma_0$ (\ref{eq:sigma0mat})
and $J$ (\ref{eq:Jdef}), it follows that:
\begin{eqnarray}
\langle Z(\varepsilon) \rangle =
\int \mathcal{D}\sigma \exp \{ 
-2 N \omega^{+^\prime} 
\trg T^{-1}_0 (\sigma_0 + N^{-1/2} \delta P) T_0 L -
\makebox[100pt]\nonumber  \\
\frac{1}{2} \trg \left[1 - 
(\sigma_0 + 2 \omega^{+^\prime} L)^2 \right](\delta P)^2 +
2 N^{1/2} \omega^{+^\prime} \trg L \delta P -
\nonumber \\
\frac{N}{\lambda} \trg 
\left[ T^{-1}_0 (\sigma_0 + N^{-1/2} \delta P) T_0 +
2 \omega^{+^\prime} L \right] J +
O(N^{-1/2}) \}. \quad
\label{eq:Zfinal}
\end{eqnarray}

The above equation makes it clear that the degrees of freedom of
$\sigma$, contained in the parametrization of $\delta P$, correspond
to the massive fluctuations around the saddle manifold. Only the
quadratic fluctuations are kept in the previous equation. The others
are suppressed by the factor $N$ to negative powers. Furthermore,
using the explicit value for $\sigma_0$ (\ref{eq:sigma0}), the
fluctuations become massless at the end of the spectrum, indicating
that the procedure fails for the energies close to the end of the
spectrum.

In the standard supersymmetry method, it is assumed that
$\omega^\prime \sim N^{-1}$ and only the highest order terms in the
exponent of the partition function (\ref{eq:Zfinal}) are kept.
Further, all terms proportional to $\omega$ in the expressions for the
saddle point (\ref{eq:sigma0}) are neglected. As a consequence, the
massive modes decouple and are integrated out. The expression for the
averaged partition function obtained by the above approximations
defines the non-linear zero-dimensional $\sigma$ model. The leading
order term of the connected part of the two point correlation
functions can be calculated from such partition function
\cite{VWZ}\cite{Zuk}.  If the corrections to that result are needed,
or the same correlator have to be calculated for $\omega^\prime \sim
1$, other terms have to be considered also, as in (\ref{eq:Zfinal})
and (\ref{eq:sigma0}). It is the interaction between the modes
parameterizing the matrix $T_0$ and the massive modes, $\delta P$ that
give rise to the non-leading order terms.

\section{The Correlation Functions}
\label{sec:cor}

The correlation functions that have to be calculated are obtained
from the second derivative of the partition function with respect to
the sources, as can be seen from (\ref{eq:CbyGG}) and
(\ref{eq:GGbydZ}). Differentiating (\ref{eq:Zfinal}) and setting the
sources to zero, the expression whose evaluation is the main aim in
this section is obtained:
\begin{equation}
I =
4 \langle {G(E_1) G(E_2)} \rangle =
\frac{ \partial^2 }{\partial \varepsilon_1 \partial \varepsilon_2}
\left. \langle Z(\varepsilon) \rangle \right|_{\varepsilon = 0} =
\int \mathcal{D}\sigma K_{11} K_{22} e^{B + F},
\label{eq:ddZ}
\end{equation}
where:
\begin{equation}
K_{pp}  =  
\frac{N}{\lambda}
\trg \left[ T^{-1}_0 (\sigma_0 + N^{-1/2} \delta P) T_0 +
2 \omega^{+^\prime} L \right]_{pp}\!\! k,
\quad (p = 1, 2), \quad 
k^{\alpha \alpha^\prime} = (-)^\alpha \delta^{\alpha \alpha^\prime}
\label{eq:kpp} 
\end{equation}
\begin{eqnarray} 
B + F  =  
-2 N \omega^{+^\prime}
\trg T^{-1}_0 (\sigma_0 + N^{-1/2} \delta P) T_0 L +
2 N^{1/2} \omega^{+^\prime} \trg L \delta P -
\nonumber \\
\frac{1}{2} \trg \left[1 - 
(\sigma_0 + 2 \omega^{+^\prime} L)^2 \right](\delta P)^2.
\label{eq:totexp}
\end{eqnarray}
The measure is given by (\ref{eq:Ds}), $B$ is the part of the
exponent that does not contain any Grassman variables and $F$ is the
Grassman dependent part of the exponent.

The calculation of the above expression begins with the evaluation of
the exponent:
\begin{eqnarray}
B  = 
-2 N \omega^{+^\prime} 
(\sigma_{01} - \sigma_{02}) (\lambda_0 - \lambda_1) -
\makebox[200pt] \nonumber \\
2 N^{1/2} \omega^{+^\prime} 
\left[ (\lambda_0 - 1) (a_1 - a_2) - i (\lambda_1 - 1) (b_1 - b_2)
\right] -
\makebox[60pt]{} \nonumber \\
\frac{1}{2} \left[ 1 - (\sigma_{01} + 2 \omega^{+^\prime})^2
\right] (a_1^2 + b_1^2) -
\frac{1}{2} \left[ 1 - (\sigma_{02} - 2 \omega^{+^\prime})^2
\right] (a_2^2 + b_2^2)
\label{eq:bosexp} 
\end{eqnarray}
\begin{equation}
F = 
2 N^{1/2} \omega^{+^\prime} (\lambda_0 - \lambda_1) 
\left[ (a_1 - i b_1) (\eta^* - \zeta_1^*) (\eta - \zeta_1) +
(a_2 - i b_2) (\rho^* - \zeta_2^*) (\rho - \zeta_2) \right].
\label{eq:fermex}
\end{equation}
It is understood that $\sigma_{0p}$ take value:
 $\sigma_{0p}^{+\pm}$
depending on which one of the two two-point correlators, $
\langle G^+(E_1) G^{\pm}(E_2) \rangle $, is calculated.

Next, the Grassmanian part of the exponent is expanded, it is
multiplied by the rest of the integrand of (\ref{eq:ddZ}) and the
product expanded in terms of the Grassman variables. The
integration over the Grassman variables is done using Wegner's theorem
\cite{Wegth}. The special case of the theorem needed here is presented
in appendix \ref{ap:Gint}. The important steps of the integration over
the Grassman variables are summarized in appendix \ref{ap:GGc}, leading to
the conclusion that among all terms in the expansion of the integrand
of (\ref{eq:ddZ}) in terms of the powers of Grassman variables, only
two contribute to the connected correlator, $\langle
{G(E_1) G(E_2)} \rangle^c$: the term with factor $\eta^* \eta \rho^*
\rho$, denoted by: $\langle {G(E_1) G(E_2)} \rangle_{(\zeta = 0)}^c$,
and the term with factor $\eta^* \eta \rho^* \rho \zeta^*_1 \zeta_1
\zeta^*_2 \zeta_2$, denoted by: $\langle {G(E_1) G(E_2)}
\rangle_{(max)}^c$:
\begin{equation}
\langle {G(E_1) G(E_2)} \rangle^c 
= \langle {G(E_1) G(E_2)} \rangle_{(\zeta = 0)}^c +
\langle {G(E_1) G(E_2)} \rangle_{(max)}^c.
\label{eq:GGc}
\end{equation}
If the appropriate terms of (\ref{eq:ddZ}) are extracted, the
integration over the Grassman variables can be completed. Using the
explicit expression for the measure, (\ref{eq:Ds}), and integrating
over $\phi_0$ and $\phi_1$, the following expression for the connected
correlator is obtained:
\begin{equation}
N^{-2} \langle {G(E_1) G(E_2)} \rangle^c_{(\zeta = 0)} = 
\frac{1}{4 \lambda^2} 
\int^{\infty}_1 d\lambda_0 \int^1_{-1} d\lambda_1 
\left( \sigma_{01} - \sigma_{02} \right)^2 
e^{-2 N \omega^\prime (\lambda_0 - \lambda_1) (\sigma_{01} -
\sigma_{02})} 
\label{eq:neta} 
\end{equation}
\begin{eqnarray}
N^{-2} \langle {G(E_1) G(E_2)} \rangle_{(max)}^c = 
\makebox[250 pt]{} \nonumber \\
\frac{-1}{4 \lambda^2 (2 \pi )^2} 
\int^{\infty}_1 d\lambda_0 \int^1_{-1} d\lambda_1 
\int^{\infty}_{-\infty} \frac{da_1 da_2 db_1 db_2 I(P) I_{max} e^B}
{(\lambda_0 - \lambda_1)^2 (a_1 - i b_1)^2 (a_2 - i b_2)^2} 
\label{eq:feta}
\end{eqnarray}
where $I(P)$ and $B$ are given in (\ref{eq:ipdef}) and
(\ref{eq:bosexp}), respectively, and:
\begin{eqnarray}
I_{max} =
(a_1 - i b_1) (a_2 - i b_2) (\lambda_0 - \lambda_1)^2 
\left\{
2 N^{-1} - 4 \omega^{+^\prime} (a_1 + a_2 - i b_1 - i b_2)^2 + 
\right. \makebox[10pt] \nonumber \\ \left.
8  \omega^{+^\prime} \lambda_1 
\left[ \sigma_{01} - \sigma_{02} + i N^{-1/2} (b_1 - b_2) \right] +
4 N \omega^{+^{\prime^2}} \lambda_1^2  
\left[ \sigma_{01} - \sigma_{02} + i N^{-1/2} (b_1 - b_2) \right]^2 -
\right. \nonumber \\ \left. 
8 N \omega^{+^{\prime^2}} \lambda_0 \lambda_1
\left[ \sigma_{01} - \sigma_{02} + N^{-1/2} (a_1 - a_2) \right] 
\left[ \sigma_{01} - \sigma_{02} + i N^{-1/2} (b_1 - b_2) \right]^2 -
\right. \quad \label{eq:intl1}
\\ \left.
8  \omega^{+^\prime} \lambda_0 
\left[ \sigma_{01} - \sigma_{02} + N^{-1/2} (a_1 - a_2) \right] +
4 N \omega^{+^{\prime^2}} \lambda_0^2 
\left[ \sigma_{01} - \sigma_{02} + N^{-1/2} (a_1 - a_2) \right]^2 
\right\}
\nonumber
\end{eqnarray}

The final step is the integration of (\ref{eq:neta}) and (\ref{eq:feta})
over the ordinary variables.  The crucial problem at this point is the
convergence of the integrals. Clearly, from (\ref{eq:bosexp}), the
integrals over $a_1,\ a_2,\ b_1$ and $b_2$ are Gaussian, therefore
convergent. Furthermore, since $\lambda_0 - \lambda_1 \geq 0$ over the
entire integration range, the integrals in (\ref{eq:neta}) and
(\ref{eq:feta}) are convergent as long as the real part of $-2 N
\omega^\prime (\sigma_{01} - \sigma_{02}) (\lambda_0 - \lambda_1)$ is
negative. The other terms in the exponent (\ref{eq:bosexp}) are
small, since they contain the massive fluctuations: $a_1, a_2, b_1,
b_2 < N^{1/2}$. Only in that case does the perturbative expansion
around the saddle point make sense. The convergence properties, as
well as the final calculations, are given separately for the two
regimes in the next two sub-sections.

\subsection{Large $\omega$ regime}

In the large $\omega$ regime ($\omega^\prime$ of the order of, but
smaller than $1$) one looks for the smoothed correlators -- all terms
oscillating with frequencies of the order $N$ are neglected. (More
precisely, the correlator is averaged over the scale few times bigger
than $N^{-1}$.) The idea is that the oscillations are due to the
discrete energy spectrum when $N$ is finite (isolated singularities of
the Green's function), which should disappear when $N \rightarrow
\infty$ (the singularities merge to form a branch cut). The
integrations over $\lambda_0$ and $\lambda_1$ in each of the
expressions for the two parts of the connected correlator  
(\ref{eq:neta}) and (\ref{eq:feta}) give four terms. The only
non-oscillating term comes from the boundary values $\lambda_0 = 1 =
\lambda_1$. Nevertheless, even the neglected oscillating terms have to
be finite in order for the procedure to be valid. In order to motivate
the forthcoming discussion, one can easily see that the connected two
point correlator (\ref{eq:GGc}), remains the same even if
some of the massive modes, $a_1, a_2, b_1$ or $b_2$, are scaled by an
arbitrary complex constant $c$, provided that the real part of the
quadratic term in the exponent remains negative. It may be surprising,
although it is easy to check, that, considering the same equations and
bearing in mind the relation between $\mu_p \bar{\mu}_p$ and
$\lambda_p$ (\ref{eq:lambdadef}), if both $\mu_0 \bar{\mu}_0$ and
$\mu_1 \bar{\mu}_1$ are scaled by $c$, or equivalently $\lambda_p
\mapsto c \lambda_p + 1 - c$, the term that survives smoothing, after
the integrations over $\lambda_0$ and $\lambda_1$ are done, does not
explicitly depend on $c$. Therefore, $c$ can be chosen such that
$\Real\ c\, \omega^+ (\sigma_0^{+\pm} - \sigma_1^{+\pm})$ is positive
infinitesimal so that all terms are convergent, while the
non-vanishing term is unaffected by the scaling. In effect, what has
been done is to rotate the contour of the integration of $\lambda_0$
and $\lambda_1$ in the complex plane, while keeping the branch-cut
structure of the Green's function intact. The rotation is justified
because the integrands of (\ref{eq:neta}) and (\ref{eq:feta}) are
polynomials in $\lambda_0$ and $\lambda_1$, so the only poles are at
infinity. Furthermore, from the discussion given in section
\ref{sec:bes}, the upper limit of $\lambda_0$ has to be understood as
large but finite. It is set to infinity only after the limit \mbox{$N
\rightarrow \infty$} is taken.

The above procedure may resemble the one used in the orthogonal
polynomials method, in which the same connected correlator was
calculated at infinity in the complex plane and then analytically
continued to the real axis, making sure that the appropriate
branch-cut is obtained \cite{Eynard}. Since the correlator in
that case is calculated at infinity, the oscillatory parts do not
exist: the function is smoothed by construction. Similarly, the
scaling by an arbitrary constant $c$ given in this article, introduces
small negative parts in the exponentials of the oscillating terms,
setting them to zero.

The following are the final expressions for the smoothed connected
correlator $\langle {G^+(E_1) G^{\mp}(E_2)}\rangle$ in the large
$\omega$ regime, obtained by the straightforward integration of
(\ref{eq:neta}) and (\ref{eq:feta}).
\begin{equation}
N^{-2} \langle {G^+(E_1) G^{\pm}(E_2)}_{(\zeta = 0)} \rangle^c  = 
\frac{1}{4 \lambda^2} \frac{1}{(2 N \omega^\prime)^2} 
\end{equation}
\begin{eqnarray}
N^{-2} \langle {G^+(E_1) G^{\pm}(E_2)} \rangle_{(max)}^c = 
\makebox[270 pt]{}
\label{eq:GGmes}  \\
\frac{1}{\lambda^2} 
\Real \left\{
\frac{1}{N^2 (\sigma_{01}^{+\pm} - \sigma_{02}^{+\pm}) 
\left[1 - (\sigma_{01}^{+\pm} + 2 \omega^\prime)^2\right]
\left[1 - (\sigma_{02}^{+\pm} - 2 \omega^\prime)^2\right]} +
\right.   \nonumber
\\ \left.
 \frac{5}{N^4} 
\frac{ \left[ 1 - (\sigma_{01} + 2 \omega^\prime)^2 \right]^2 -
6 \left[ 1 - (\sigma_{01} + 2 \omega^\prime)^2 \right] 
\left[ 1 - (\sigma_{02} - 2 \omega^\prime)^2 \right] +
\left[ 1 - (\sigma_{02} - 2 \omega^\prime)^2 \right]^2}
{(\sigma_{01} - \sigma_{02})^6 
\left[ 1 - (\sigma_{01} + 2 \omega^\prime)^2 \right]^3 
\left[ 1 - (\sigma_{02} - 2 \omega^\prime)^2 \right]^3} 
\right\} 
\nonumber
\end{eqnarray}
In order to derive the above relations, the denominator of $I(P)$,
(\ref{eq:ipdef}), has been expanded in terms of the power series. The
validity of the expansion is based on the fact that only the small
quadratic fluctuations are considered, that is, the massive modes are
smaller than $N^{1/2}$ over the entire integration range.

Using (\ref{eq:CbyGG}) and (\ref{eq:sigma0}), the smoothed
ensemble-averaged energy density-density correlation function, in the
large $N$ limit and for: $E_1 - E_2 \sim N^0$, provided the
energies are not close to the endpoints of the spectrum, is found to
be:
\begin{eqnarray}
C(E_1, E_2) = 
\frac{-1}{2 \pi^2 N^2} 
\frac{4 \lambda^2 - E_1 E_2}{(E_1 - E_2)^2 
\sqrt{4 \lambda^2 - E_1^2} \sqrt{4 \lambda^2 - E_2^2}} +
\makebox[100pt] \nonumber \\
\frac{40 \lambda^6}{N^4 \omega^6}
\frac{U^2 (S U - T) - (\frac{\omega}{4 \lambda})^2 (3 S U - T)}
{(\sqrt{1 - (E_1/2 \lambda)^2} \sqrt{1 - (E_2/2 \lambda)^2})^3}, 
\label{eq:Cfin}
\end{eqnarray}
where:
\begin{equation}
S  = 
-4 (4 \lambda^2 - E_1^2) (4 \lambda^2 - 2 E_1^2) -
4 (4 \lambda^2 - E_2^2) (4 \lambda^2 - 2 E_2^2) +
24 4 (4 \lambda^2 - E_1^2) (4 \lambda^2 - E_2^2)
\end{equation}
\begin{eqnarray}
T & = &
24 (2 \lambda)^2 (4 \lambda - E_1^2) (4 \lambda - E_2^2) E_1 E_2 
\\
U & = & 
4 \lambda^2 - E_1 E_2.
\end{eqnarray}

It should be noted that the usual supersymmetry method
(zero-dimensional $\sigma$ model) does not give a result in this regime.
The leading order of the above result is identical to the result
obtained by Br\'{e}zin and Zee \cite{BZ} giving another confirmation
of the correctness of the above stated convergence prescription. As
for the sub-leading term, it is questionable if any of the neglected
parts, such as higher than quadratic fluctuations, may give
contribution of the same order. Nevertheless, it is correct within the
current approximation, namely small quadratic fluctuations around the
saddle point.

\subsection{Small $\omega$ regime}

Let us consider the convergence of $\langle {G^+(E_1) G^-(E_2)}$
first.  Again, the convergence is achieved if the real part of $2 N
\omega^\prime (\sigma_{01} - \sigma_{02}) (\lambda_0 - \lambda_1)$ is
positive.  From (\ref{eq:sigma0}), for small positive $\epsilon$ and
for small but finite $\omega^\prime$ (of the order $N^{-1}$), it
follows that:
\begin{equation}
\Real \omega^{+^\prime} (\sigma_{01}^{+-} - \sigma_{02}^{+-}) =
-2 \omega^{\prime^2} 
\left(1 + \frac{\epsilon}{(1 - E^{\prime^2})^{3/2}}\right) +
2 \epsilon (1 - E^{\prime^2})^{1/2}.
\end{equation} 

Therefore the convergence can be achieved by taking the $N \rightarrow
\infty$ limit first, so that $\omega \rightarrow 0$ while $\epsilon$
is still positive, making the above expression positive. Only then
can $\epsilon$ be set to zero giving the required correlator.
Using the procedure just stated, the terms coming from the upper bound 
of the integration over $\lambda_0$ in (\ref{eq:neta}) and (\ref{eq:feta})
vanish after the limit $N \rightarrow \infty$ is taken. The
integration over $\lambda_1$ gives the non-exponential terms (from
$\lambda_1 = 1$) and the purely oscillatory terms ($\lambda_1 = 1$),
rendering the integration convergent.  Note that in the usual
supersymmetry method ($\sigma$-model) the terms proportional to
$\omega$ in the expressions for $\sigma_0$, (\ref{eq:sigma0}), are
neglected and consequently the integrals are immediately
convergent. The above analysis was necessary for the calculations
beyond the $\sigma $-model, but could also be thought of as a
clarification of the usual procedure.

The integrations needed in (\ref{eq:neta}) and (\ref{eq:feta})
are now straightforward, with the following result:
\begin{equation}
N^{-2} \langle {G^+(E_1) G^{-}(E_2)} \rangle_{(\zeta = 0)}^c = 
\frac{1}{4 \lambda^2} \frac{1}{(2 N \omega^\prime)^2} 
\left[ 1 - e^{8 i N \omega^\prime \sqrt{1 - E^{\prime 2}}} \right]
\label{eq:smomze}
\end{equation}
\begin{eqnarray}
N^{-2} \langle {G^+(E_1) G^{-}(E_2)} \rangle_{(max)}^c = 
\makebox[250 pt] \nonumber \\
\frac{-1}{4 \lambda^2} \frac{1}{(2 \pi)^2}
\left\{
\frac{-16 \pi^2}{N^2 (\sigma_{01}^{+-} - \sigma_{02}^{+-}) 
\left[1 - (\sigma_{01}^{+-} + 2 \omega^\prime)^2\right]
\left[1 - (\sigma_{02}^{+-} - 2 \omega^\prime)^2\right]} +
\right. \nonumber \\ \left. 
\left[  
\frac{-16 \pi^2}{N^2 (\sigma_{01}^{+-} - \sigma_{02}^{+-}) 
\left[1 - (\sigma_{01}^{+-} + 2 \omega^\prime)^2\right]
\left[1 - (\sigma_{02}^{+-} - 2 \omega^\prime)^2\right]} +
\right. \right. \\ \left. \left. 
\frac{(8 \pi \omega^\prime)^2}
{\left[1 - (\sigma_{01}^{+-} + 2 \omega^\prime)^2\right]
\left[1 - (\sigma_{02}^{+-} - 2 \omega^\prime)^2\right]}
\right] 
e^{8 i N \omega^\prime \sqrt{1 - E^{\prime 2}}}
\right\} \nonumber
\end{eqnarray}

In order to derive the above relations, the denominator of $I(P)$,
(\ref{eq:ipdef}), as well as the factor $\exp \{-4 i N^{1/2}
\omega^\prime (b_1 - b_2)\}$ appearing after the $\lambda_1$ integration
in (\ref{eq:feta}), have been expanded in terms of the power
series. The validity of the expansions is based on the fact that only
the small quadratic fluctuations are considered, that is, the massive
modes are smaller than $N^{1/2}$ over the entire integration range.

Using the expressions for the saddle points (\ref{eq:sigma0}) the
final expression is obtained:
\begin{eqnarray}
N^{-2} \langle {G^+(E_1) G^{-}(E_2)} \rangle_{(max)}^c = 
\makebox[250 pt] \nonumber \\
\frac{-1}{16 \lambda^2 N^2 (1 - E^{\prime^2})}
\left\{ \frac{1}{1 - E^{\prime^2}} -
\left[ \frac{1}{1 - E^{\prime^2}} + 16 \omega^{\prime^2} N^2 \right]
e^{8 i N \omega^\prime \sqrt{1 - E^{\prime^2}}} \right\}.
\label{eq:smomax}
\end{eqnarray}

The equations (\ref{eq:smomze}) and (\ref{eq:smomax}) together constitute
the final result for the connected two point correlation function $
\langle G^+(E_1) G^-(E_2) \rangle $ in the small $\omega$
regime. While the former is a well known result obtained using the
standard supersymmetry method, among other methods, the latter is the
leading order correction to the result.

\section{Conclusion}

One of the methods used in RMT is the supersymmetry method. The usual
procedure includes some approximations and leads to the zero
dimensional $\sigma$-model. In this article, the Gaussian unitary
ensemble is considered and the supersymmetry method is extended to
keep some of the previously neglected terms. The partition function
for the extended method is given in (\ref{eq:Zfinal}). It contains
the $\sigma$-model terms, as well as the lower order terms (quadratic
fluctuations around it).

In the large energy difference regime, the extended supersymmetry
method is used to calculate the connected energy density-density correlation
function (\ref{eq:CbyGG}). The usual degeneracy of the saddle point is lifted
fully. Using a symmetry of the model (invariance under rescaling of
the variables $\mu_p \bar{\mu}_p$) the potentially divergent integrals
can be made finite. As a consequence the smoothed connected energy
density-density correlation function is calculated,
(\ref{eq:Cfin}). Both the leading order term, which agrees with the
known result \cite{BZ} obtained by the method of the orthogonal
polynomials, and the highest order corrections in the $N^{-1}$
expansion are contained in the
result. Furthermore, both contributions can be seen to come from terms
that are neglected in the usual supersymmetry approach.

As for the small energy difference regime, the degeneracy of the
saddle point is only slight. The contributions from the
(non-degenerate) saddle point, as well as from the quadratic
fluctuations are taken into account. Convergence can be achieved
provided that the right limiting procedure is used. As the result, the
leading correction to the well known result for \mbox{$\langle
G^+(E_1) G^-(E_2) \rangle^c $} is calculated (\ref{eq:smomax}).

A natural extension of the work presented here would be to consider
other ensembles (orthogonal, symplectic, $\ldots$). Also, the
application of the modified supersymmetry method to non-Gaussian
ensembles seems worth investigating.


\section*{Acknowledgements}

The author would like to thank Y. Srivastava for useful discussions.


\appendix

\section{Grassman algebra: definitions and conventions}
\label{ap:Gdef}

Grassman variables, $\eta_1$ and $\eta_2$ are anti-commuting: 
\begin{equation}
\eta_1 \eta_2 = - \eta_2 \eta_1.
\end{equation}
The complex-conjugate of a Grassman variable is defined as:
\begin{equation}
(\eta_1 \eta_2)^* = \eta_1^* \eta_2^*, \quad
\eta^{**} = -\eta.
\end{equation}

A general super-vector $\varphi$ is defined as:
\begin{equation}
\varphi = (z_1,\ \eta_1,\ z_2,\ \eta_2,\ \ldots )
\end{equation}
where $z_p$ and $\eta_p$ are vectors with ordinary and Grassman
elements respectively.  The supervector $\varphi$ used in this article
((\ref{eq:L1}) and below) has the form: 
\begin{equation}
\varphi = (\varphi^0_1,\ \varphi^1_1,
\ \varphi^0_2,\ \varphi^1_2), 
\end{equation}
where $\varphi^{\alpha}_p$ is a $N$-dimensional vector consisting of
ordinary ($\alpha = 0$), or Grassman ($\alpha = 1$) elements
$\varphi^{\alpha}_p(\mu)$ with $\mu = (1, \ldots , N)$. Also,
$\bar{\varphi}$ is defined as $\bar{\varphi} =
\varphi^\dagger s$ where $s = (1_N, 1_N, -1_N, 1_N)$, or explicitly:
\begin{equation}
\overline{\varphi} = ({\varphi^0_1}^*, {\varphi^1_1}^*, -{\varphi^0_2}^*, {\varphi^1_2}^*). 
\end{equation}
  
If $a$ and $b$ are ordinary (Bosonic) and $\alpha$ and $\beta$ are
Grassman (Fermionic) 
variables or matrices, the super-transpose and Hermitian adjoint of a
super-matrix $A$ are:
\begin{equation}
A = \left( \begin{array}{cc}
	a & \alpha \\
	\beta & b \end{array} \right),\quad
A^T = \left( \begin{array}{cc}
	a^T & \beta^T \\
	-\alpha^T & b^T \end{array} \right)\quad
\mbox{and}\quad
A^\dagger = A^{*T}.
\end{equation}
$a$, $b$, $\alpha$ and $\beta$ are called boson-boson,
fermion-fermion, boson-fermion and fermion-boson blocks.  The
super-matrix $A$ was given in what it is called a boson-fermion
notation. A super-matrix in a general form can be obtained from a
super-matrix in the boson-fermion block notation by interchanging some
of its rows and columns, so that it appears to consist of blocks,
each block having the boson-fermion block form. The definitions for
the super-transpose and Hermitian adjoint are to be applied by putting
the matrix in boson-fermion notation first and then using the above
definitions. 

Elements, or blocks, of both super-vectors and super-matrices used in
this article have their superscripts denoting grading ($0$ for
ordinary and $1$ for Grassman variables), while the subscripts denote
blocks, each having the boson-fermion block form.

\section{Grassman integration and Wegner's theorem}
\label{ap:Gint}

In this appendix, the integration over the Grassman variables is defined, 
the Gaussian super-integral is given and a special case of Wegner's
integral theorem, needed in this article, is presented. 

Grassmanian integration is defined as:
\begin{equation}
\int d\eta = 0, \qquad \int d\eta \eta = (2 \pi)^{-1/2}
\end{equation}
It is easy to extend Gaussian integration to include both ordinary and
Grassman variables:
\begin{equation}
\int^{\infty}_{-\infty} \prod_{k=1}^{N} i dz_k dz^*_k 
\int \prod_{k=1}^{N} d\eta_k d\eta^*_k e^-\phi^\dagger A \phi
= \mbox{detg}^{-1} A,
\label{eq:superGauss}
\end{equation}
where $\phi = (z, \eta)$ is a $2N$-dimensional supervector, $A$ is a
super-matrix in the boson-fermion notation and detg is the graded
(super) determinant \cite{Nath}. The above integral is commonly
referred to as the Gaussian superintegral.

A property of the Grassman integration, called Wegner's integral
theorem, deserves clarification. Although it is not often
explained, it is important for the calculations in this article.  The
motivation for the theorem comes from considering the following simple
Gaussian superintegral:
\begin{equation}
I \equiv 
\int dp dq d\eta d\eta^* 
e^{-\frac{1}{2} (p^2 + q^2) + \eta \eta^*} = 1
\label{eq:Wegidentity}
\end{equation}
On the other hand, performing the following change of variables:
\begin{eqnarray}
p = u + \zeta^* \zeta (u - i v), \qquad
\eta = \zeta^* (u - i v),
\nonumber \\
i q = i v - \zeta^* \zeta (u - i v), \qquad
\eta^* = \zeta (u - i v)
\end{eqnarray}
and including the Jacobian factor: $(u - i v)^{-2}$, the integral
becomes:
\begin{equation}
I = 
\int du dv d\zeta d\zeta^* (u - i v)^{-2} 
e^{-\frac{1}{2} (u^2 + v^2)}.
\label{eq:Wegintegral}
\end{equation}
Although there is no Grassman dependence in the integrand, the
integral is not zero because the integration over the ordinary
variables is singular. The correct result was already given in
(\ref{eq:Wegidentity}). In many cases, when an integral of the form
of (\ref{eq:Wegintegral}) is encountered, it is neither easy, nor
convenient to transform it to the form of
(\ref{eq:Wegidentity}). The theorem gives the prescription on how to
evaluate such an integral. Instead of the full theorem, only a special
case of the theorem, relevant for the calculations presented in this
article, is now presented. A more precise formulation is given in
\cite{Wegth}.

Consider an integral of the form:
\begin{equation}
I = \int \mathcal{D} \mu(Q) f(Q),
\end{equation}
where $Q$ is a super-matrix that can be parametrized as a coset space:
$Q = U^{-1} \Lambda U$. $U$ contains Grassman and $\Lambda$ contains
ordinary variables, while $\mathcal{D} \mu(Q)$ has a singularity at the
origin of the coset space. $f$ can be expanded as $f(Q) = f_0 +
f_{max} + f_{other}$, where $f_0$ and $f_{max}$ are
the terms in the expansion of $f(Q)$ in the powers of Grassman
variables that contain no Grassman variables and the product of
all other terms in the expansion. If $f_{other}$ goes to zero
at the origin, the integral $I$ can be evaluated
to give:
\begin{equation}
I = \left. \int \mathcal{D} \mu(Q) f_{max} + f_{0}\right|_{\Lambda =
\Lambda_0},
\end{equation}
where $\Lambda_0$ is the origin of the coset space.

\section{Integration Over Grassman Variables}
\label{ap:GGc}

The integration over Grassman variables in the integral (\ref{eq:ddZ})
is outlined in this appendix. As a result, the terms in the Grassman
expansion of the integrand relevant for the connected correlators,
(\ref{eq:CbyGG}), are identified.
   
The integration over the Grassman variables of the matrix $T_0$ ($\eta,\ 
\eta^*,\ \rho\ \mbox{and}\ \rho^*$) can be done by expanding the
integrand in terms of those variables using the obvious notation: 
\begin{equation}
I = \int \mathcal{D}\sigma (f_0 + f_{max}  \eta^* \eta \rho^*
\rho + 
f_{other}),
\label{eq:etaint}
\end{equation}
$f_{other}$ vanishes at the origin of the $T_0$ coset
$\lambda_0 = 1 = \lambda_1$ (the $\phi_{0,1}$ are not important because
they decouple). The Wegner's theorem applies, giving:
\begin{equation}
I = \left. \int \mathcal{D} P I(P) f_0\right|_{T_0 = 1} + 
\int \mathcal{D}\sigma f_{max}. 
\label{eq:I}
\end{equation}

The first term of the above equation can be easily computed using
(\ref{eq:ddZ}), (\ref{eq:kpp}) and (\ref{eq:totexp}):
\begin{equation}
\left. \int \mathcal{D} P I(P) f_0\right|_{T_0 = 1} =
\frac{4 N^2}{\lambda^2} (\sigma_{01} + 2 \omega^{+^\prime})
(\sigma_{02} - 2 \omega^{+^\prime}).
\end{equation}
On the other hand, $\langle {G(E_p)} \rangle $ can be calculated
starting from (\ref{eq:GbydZ}) and using (\ref{eq:Zfinal}),
(\ref{eq:kpp}) and (\ref{eq:totexp}). Wegner's theorem is used again to
show that the only contribution comes from the term that has no
Grassman variables and gives:
\begin{equation}
\langle {G(E_p)} \rangle =
-\frac{N}{\lambda} (\sigma_{0p} - (-)^p 2 \omega^{+^\prime}).
\end{equation}

Therefore, the first term in (\ref{eq:I}) (divided by 4) is exactly
the disconnected part of the correlator $\langle {G(E_1)}
\rangle \langle {G(E_2)} \rangle $. Therefore the connected part
(\ref{eq:Gcdef}), is simply:
\begin{equation}
\langle {G(E_1) G(E_2)} \rangle^c = 
\frac{1}{4} \int \mathcal{D}\sigma f_{max} \equiv
\frac{1}{4} I^c.
\end{equation}

Next, $I^c$ is integrated over the Grassman variables contained in
$\delta P$: $\zeta_1,\ \zeta_1^*$, $\zeta_2$, and $\zeta_2^*$. The
procedure is the same as above, only this time it has to be done
separately for $\delta P_1$ and $\delta P_2$ because they are
independent and both of them have a coset structure. The origins of
the respective coset spaces are given by: $a_1 = 0 = b_1$ and $a_2 = 0
= b_2$, so that $I^c$ can be expanded as:
\begin{equation}
I^c = \int \mathcal{D} \sigma \left[ {g_p}_0 + 
{g_p}_{max}  \zeta^*_p \zeta_p + {g_p}_{other} \right],
\end{equation}
in obvious notation, for both $p = 1$ and $p= 2$. Both ${g_p}_{other}$
vanish at the respective origins, so Wegner's theorem
applies. Consequently, the connected correlation function receives only
two contributions: one from the term with factor $\eta^* \eta \rho^*
\rho$ evaluated at the origin of $\delta P$ coset (denoted by
subscript $(\zeta = 0)$ ) and the other from the term with all
Grassman variables (denoted by subscript $(max)$). Therefore the above
statement can be written as:
\begin{equation}
\langle {G(E_1) G(E_2)} \rangle^c =  
\langle {G(E_1) G(E_2)} \rangle^c_{(\zeta = 0)}  +
\langle {G(E_1) G(E_2)} \rangle^c_{(max)}.
\end{equation}

The above equation specifies how to perform the integration over the
Grassman variables needed in section \ref{sec:cor}.

\end{document}